\documentclass[preprint,aps]{revtex4}
\begin{document}
\title{Quantifying energy condition violations in traversable wormholes}
\author{Sayan Kar}
\email{sayan@cts.iitkgp.ernet.in}
\affiliation{Department of Physics and Centre for Theoretical Studies
\\
Indian Institute of Technology, Kharagpur 721 302, WB, India}
\author{Naresh Dadhich}
\email{nkd@iucaa.ernet.in}
\affiliation{Inter--University Centre for Astronomy and Astrophysics \\
Post Bag 4, Ganeshkhind, Pune 411 007, India}
\author{Matt Visser}
\email{matt.visser@vuw.ac.nz}
\homepage{http://www.mcs.vuw.ac.nz/~visser}
\affiliation{School of Mathematical and Computing Sciences,
Victoria University of Wellington, New Zealand}

\begin{abstract}
The `theoretical' existence of traversable Lorentzian wormholes 
in the classical, macroscopic world is plagued by the violation of
the well--known energy conditions of General Relativity. In this brief
article we show : (i) how the extent of violation can be quantified
using certain volume integrals (ii) whether this `amount of
violation' can be minimised for some  specific cut--and--paste geometric
constructions. Examples and possibilities are also outlined.    
\end{abstract}

\maketitle
\def\d{{\mathrm{d}}}
\def\be{\begin{equation}}
\def\ee{\end{equation}}

\section{The problem and our attitude}

It is well--known by now that the `theoretical' 
existence of traversable Lorentzian wormholes is plagued by the
violation of the energy conditions of General Relativity {\cite{ecviolation,book}}
. Researchers
have come up with a variety of proposals, most of which gain support
from the fact that quantum expectation values of the stress energy
tensor can often become negative {\cite{egy,hochberg}}. The experimentally verified
case of the Casimir effect {\cite{casimir}} is often cited as a `proof' of the 
existence of `matter' with `negative energy density' though in
experiments on the Casimir effect the quantity measured is the
force (and hence the pressure) of the `fluctuating vacuum' 
between parallel metallic plates. 

Leaving aside the question about whether wormholes exist or whether
negative energy is justifiable we prefer to adopt a somewhat
braver attitude based on some recent results in other areas of
physics. For instance, before it was actually seen in the laboratory
one never believed that  `negative group velocity' {\cite{negvg}} or `negative
refractive index' {\cite{negn}} could be real. Theoretically however, these 
esoteric concepts were outlined decades ago and largely forgotten.
The same also holds good for the Casimir effect. 
It is true that today, negative energy or wormholes are
esoteric ideas. But, following the abovementioned realisation of
negative $v_g$ and $n$ it may not be too outrageous to say that
exotic things of today might be reality (in some now--inconceivable form) 
tomorrow.
Another example in case is `dark energy' which seems to dominate
70 percent of the matter in the Universe today {\cite{de}}. Dark energy has
`negative' pressure which is indeed counterintuitive but largely
in vogue amongst today's cosmolgists. Furthermore, 
particle theorists seem to be happy with a negative
cosmological constant {\cite{rs}} which helps them solve the so--called heirarchy
problem. So, why not wormholes with `negative energy'?

Of course, negative energy or negative energy density is problematic.
But then, one must ask the question `how much negative energy'--- or
is there a way to quantify the amount of violation? There have been
attempts at such quantification through the so--called `quantum inequalities'
which are essentially similar to the energy--time uncertainty relations {\cite{quantum}}. Here we propose 
a quantifier in terms of a spatial volume integral {\cite{vkd}}. 
Using this we can show that certain `cut--and--paste
constructions' allow us to reduce this `amount of violation' to
arbitrarily small values. We provide an example of such a construction 
and conclude with some open questions {\cite{vkd}}.

\section{The energy conditions and the `volume integral quantifier'}

Let us begin by discussing some of the energy conditions in the literature.
We can classify them as `local' and `global' conditions. Among local
conditions we have the Weak Energy Condition (WEC) and the Null Energy
Condition which are stated as (for a diagonal energy momentum tensor
with energy density $\rho$ and pressures $p_i$ (i=1,2,3)) : 

\begin{equation}
\rho \ge 0,\quad \rho+p_i \ge 0 \quad (WEC) \hspace{.1in} ; \hspace{.1in}
\rho + p_i \ge 0 \quad (NEC)
\end{equation} 

Other local conditions include the Strong Energy Condition (SEC) and
the Dominant Energy Condition (DEC) (for a discussion on these see
{\cite{book}}). On the other hand, global conditions involve line integrals
along complete null or timelike geodesics and therefore yield numbers.
For example, the Averaged Null Energy Condition (ANEC) is given as: 

\begin{equation}
\int_{\lambda_1}^{\lambda_2} T_{ij}k^i k^j d\lambda \ge 0
\end{equation}
where $k^i$ is the tangent vector along a null geodesic and $\lambda$ is
the affine parameter labeling points on the geodesic.  
A useful discussion on the violation of the local and global energy
conditions in the context of both classical (exotic) or quantum stress--energy
can be found in {\cite{book,barcelo}}.

Now consider a static spherically symmetric spacetime with a line element 
given by :

\be
\d s^2 = 
- \exp[2\phi(r)] \;\d t^2 
+ {\d r^2\over1-b(r)/r}
+ r^2 (\d \theta^2 + \sin^2\theta\;\d \varphi^2).
\ee

Using the Einstein field equations, the components
of the diagonal energy--momentum tensor in an orthonormal basis 
turn out to be (units --- $G=c=1$)~\cite{book} :
\begin{equation}
\rho = {1\over8\pi} {b'\over r^2} \hspace{.1in} ;\hspace{.1in} 
p_r = {1\over8\pi} 
\left[ 
-{b\over r^3} + 2 \left\{ 1 - {b\over r} \right\} {\phi'\over r}
\right]
\end{equation}
\begin{equation}
p_t = {1\over8\pi}\left[ \left\{ 1 - {b\over r} \right\}
\left [ \phi''   + \phi' \left( \phi' + {1\over r} \right)\right]
 -
{1\over2} \left({b\over r}\right)'
\left( \phi' + {1\over r} \right)\right ]
\end{equation}
where $\rho$, $p_r$, and $p_t$ are the energy density, the radial and
tangential pressures respectively.  

The ANEC integral along a radial null geodesic is 
\begin{eqnarray} 
I= 
\oint
[\rho+p_r] \exp[-2\phi]\;\d \lambda 
= \oint [\rho+p_r] \exp[-\phi]\;\d\eta \nonumber \\ 
=-\frac{1}{4\pi}\oint \frac{2}{r} \frac{\d}{\d r}\left [ e^{-\phi} \sqrt{1-\frac{b(r)}{r}}\right ] \d r \nonumber \\
=- {1\over4\pi} \oint {1\over r^2} e^{-\phi} \sqrt{1-{b\over r}} \; \d r
< 0.  
\end{eqnarray}
where $\eta$ is the proper radial distance and we have performed an
integration by parts in the last step. Both the local and averaged
energy conditions are violated by wormholes {\cite{ecviolation}}. A
quick way to see this is to note that for light rays the wormhole
throat behaves like a diverging lens --- light rays are not focused
and therefore there must be a violation of the null convergence
condition (or null energy condition, via the Einstein equations).

However, the ANEC is a line integral and therefore not very helpful 
for quantifying the `amount of violation'. This prompts us to
propose a `volume integral quantifier' which  
amounts to calculating the following definite integrals (for the
relevant coordinate domains):

\begin{equation}
\int \rho \hspace{.05in} dV \hspace{.1in} ; \hspace{.1in} \int \left [ \rho+p_i \right ] \hspace{.05in} dV
\end{equation}
with an appropriate choice of the integration measure ($4\pi r^2 \d r$ or
$\sqrt{g} \d r \d \theta \d \phi$). We define the amount of violation as the
extent to which these integrals can become negative. The important point
which we shall demonstrate below is that even if the ANEC yields a
constant negative number the volume integrals can be adjusted to 
become vanishingly small by appropriate choice of parameters.

Let us now focus on one such volume integral. Using the Einstein field
equations it is easy to check that : 
\be
\rho+p_r = {1\over8\pi r}  \left\{ 1 - {b\over r} \right\} 
\left[ \ln\left({\exp[2\phi]\over1-b/r}\right) \right]'.
\ee
Then integrating by parts
\begin{eqnarray}
&&\oint [\rho+p_r] \;\d V  = 
\left[ (r-b) \ln\left({\exp[2\phi]\over1-b/r}\right) \right]_{r_0}^\infty
\nonumber
\\
&&
\qquad\qquad
- \int_{r_0}^\infty (1-b') 
\left[ \ln\left({\exp[2\phi]\over1-b/r}\right) \right] \d r. 
\end{eqnarray}
The boundary term at $r_0$ vanishes by our construction (recall that
for a wormhole $b(r=r_0)=r_0$ where $r_0$ is the throat radius and hence the
minimum value of $r$). The boundary term at infinity vanishes because of the
assumed condition of asymptotic flatness.  Then
\be
\oint [\rho+p_r] \;\d V  = 
- \int_{r_0}^\infty (1-b') 
\left[ \ln\left({\exp[2\phi]\over1-b/r}\right) \right] \d r.
\label{E:key}
\ee
The value of this volume-integral provides information about the ``total
amount'' of ANEC violating matter in the spacetime.
One should also calculate the other volume integrals though in most cases
they do not provide any further information on the amount of violation. 

\section{An explicit `cut and paste' example}

Let us now look at a specific example.
If we consider a line element for which the spatial metric is exactly
Schwarzschild, that is $b(r)\to 2m= r_0$.  Then $\rho=0$
throughout the spacetime and we simply get :
\be
\oint p_r \;\d V  = 
- \int_{r_0}^\infty \ln\left[{\exp[2\phi]\over1-2m/r}\right] \;\d r.
\ee

Now assume that we have a wormhole whose field
only deviates from Schwarzschild ($g_{00}\neq -\left (1-2m/r\right )$)
in the region from the throat out to
radius $a>2m$. At $r=a$ we join this geometry to a Schwarzschild.
We must take care of the matching conditions --- details on these
are available in {\cite{vkd}}.  
It turns out that for this case, 
we can further simplify the above volume integral to

\be
\oint p_r \;\d V  = 
- \int_{r_0}^a \ln\left[{\exp[2\phi]\over1-2m/r}\right]\;\d r.
\ee
Under this same restriction the ANEC integral satisfies
\be
I <
- {2\over4\pi} 
\int_{a}^\infty {1\over r^2} \; \d r =  -{1\over2\pi\; a},
\ee
which is strictly bounded away from zero. (Note that while evaluating
the above from equation (5) one has to be careful about the derivative
discontinuity of $e^{-\phi}\sqrt{1-b(r)/r}$ at $r=a$. 
As discussed below,
this formula can safely be applied as $a$ approaches $2m$ from the
$a>2m$ side).
Now 
\be
\int_{r_0}^a \ln\left[{\exp[2\phi]\over1-2m/r}\right] \d r <
\int_{r_0}^a \ln\left[{\exp[2\phi_{\mathrm{max}}]\over1-2m/r}\right] \d r.
\ee
Evaluating this last integral 
\be \oint p_r \;\d V >
-(a-2m)\ln\left[{\exp[2\phi_{\mathrm{max}}]\over1-2m/a}\right] 
- 2\;m\;\ln\left({a\over2m}\right).  \ee 
This is useful because it is an explicit \emph{lower} bound on the
total amount of radial stress in terms of $\phi_{\mathrm{max}}$ and
the size of the region of ANEC violating matter. Similarly
\be \oint p_r \;\d V <
-(a-2m)\ln\left[{\exp[2\phi_{\mathrm{min}}]\over1-2m/a}\right] 
- 2\;m\;\ln\left({a\over2m}\right).  
\ee 
This is now an \emph{upper} bound in terms of $\phi_{\mathrm{min}}$
and the size of the region of ANEC violating matter.  If we now choose
geometries such that $\phi_{\mathrm{max}}$ and $\phi_{\mathrm{min}}$
are not excessively divergent, [no worse than $(a-2m)^{-\delta}$ with
$\delta<1$], we can take the limit $a\to2m^{+}$ (the superscript $+$
here means that a approaches $2m$ from the $a>2m$ side) to obtain
\be 
\oint p_r \;\d V \to 0.  
\ee
We emphasize here that the ANEC integral {\em does not} go to zero as
$a\rightarrow 2m^{+}$.

Furthermore, by considering a sequence of traversable wormholes with
suitably chosen $a$ and $\phi(r)$ [and $b(r)=2m$] we can construct
traversable wormholes with arbitrarily small quantities of
ANEC-violating matter. (With the ANEC line integral nevertheless
remaining finite and negative.)  More examples are available in
{\cite{vkd}}.

\section{Remarks and Conclusions}

The above discussion shows that we are able to (i) quantify the
amount of violation (ii) construct examples for which the violation
can be made very small. It is worthwhile to note that 
in both these constructions the local and averaged energy conditions
still remain violated and that violation cannot be made to vanish!
It might seem therefore that we have skirted the real issue by
`redefining' the notion of violation through these volume integrals.
Therefore, to prove our point we must try to establish the fact that
these volume integrals are the correct quantifiers (on physical grounds
they do seem to be so) and all theorems
which assume the validity of the averaged conditions can now be
extended to include these volume averaged conditions.  This is a task
for the future.

Finally let us place our result in the context of the four great
results of classical general relativity --- the area increase theorem
{\cite{hell,flux}}, the singularity theorem {\cite{hell}}, 
the positive mass theorem {\cite{yau}} and the topological
censorship theorem {\cite{topological}}. Each of these theorems do assume some form of an
energy condition --- the question is --- if violations are small can the
conclusions of these theorems be evaded ? It is worth noting at this
point that the conclusions of the area increase and topological censorship
thoerems can indeed be reversed by quantum--induced violations. In the 
case of the other two theorems the consequences of 
such microscopic violations may not lead to any drastic changes in their
conclusions.

\section*{Acknowledgements}

SK and ND thank the organisers of ICGC04 (Cochin) for the excellent 
hospitality provided during the Conference. SK also acknowledges
financial support from IIT Kharagpur, India.  


\end{document}